\documentclass[11pt,a4paper]{article}
\usepackage{amsmath,amssymb}
\usepackage{epsfig,graphicx}
\usepackage{cite}

\begin{document}

\title{\bf Neutrino-triggered asymmetric magnetorotational\newline mechanism for pulsar natal kick}

\author{A.~V.~Kuznetsov$^a$\footnote{{\bf e-mail}: avkuzn@uniyar.ac.ru},
N.~V.~Mikheev$^{a}$\footnote{{\bf e-mail}: mikheev@uniyar.ac.ru}
\\
$^a$ \small{\em Yaroslavl State (P.G.~Demidov) University} \\
\small{\em Sovietskaya 14, 150000 Yaroslavl, Russian Federation}
}
\date{}

\maketitle

\begin{abstract}
The sterile neutrino mechanisms for natal neutron star kicks are reanalyzed. It is shown that the magnetic field strengths needed for obtaining the observable values of kicks were underestimated essentially. Another mechanism with standard neutrinos is discussed where the outgoing neutrino flux in a supernova explosion with a strong toroidal magnetic field generation causes the field redistribution in ``upper'' and ``lower'' hemispheres of the supernova envelope. The resulting magnetic field pressure asymmetry causes the pulsar natal kick.
\end{abstract}


\section{Pulsar proper motion problem}


\def\D{\mathrm{d}} 
\def\E{\mathrm{e}}
\def\I{\mathrm{i}}


\indent\indent
The problem of large proper velocities of pulsars, born in supernova explosions (pulsar kick), 
has been discussed for more than 40 years. The total list of publications with the observational data 
is very long. We indicate here only the first papers~\cite{Shklovsky:1969,Gunn:1970} where the problem was put, 
and the papers where the data were summarized with the samples of 99 pulsars~\cite{Lyne:1994} and 
of 233 pulsars~\cite{Hobbs:2005}.  
Average speed for the sample of 233 pulsars~\cite{Hobbs:2005} was estimated at the level of 400 km/s, 
with more than 15 \% having velocities
greater than 1000 km/s. The two fastest pulsars PSRs B2011+38 and B2224+64 have {$\sim$ 1600 km/s.}

It is important that a correlation was finally established between the directions of pulsar velocities 
and of rotation axes. Initially, a conclusion was made in the paper~\cite{Deshpande:1999}, based on an 
analysis of the set of 29 pulsars, that mechanisms predicting a correlation between the rotation
axis and the pulsar velocity were ruled out by the observations. However, in the paper~\cite{Johnston:2005} 
strong observational evidence was presented for a relationship between the direction of a pulsar's motion and its rotation axis. Analysing a set of 25 pulsars which are younger than the ones taken in Ref.~\cite{Deshpande:1999}, 
the authors~\cite{Johnston:2005} conclude that 10 pulsars show an offset between the velocity vector 
and the rotation axis, which is either less than 10$^0$ or more than 80$^0$, a fraction that is very unlikely 
by random chance. 

Obviously, the reason for the initial kick is a kind of an asymmetry in a supernova explosion, but the nature of it has not yet disclosed. There were many attempts to explain this asymmetry. 

Numerous attempts to describe the effect in the hydrodynamics of a supernova explosion do not explain the large speeds. Three-dimensional simulation of the explosion with the assumption of initial asymmetry in the supernova core before the collapse, which increases during the collapse, leads to the velocity of a pulsar not more than 200 km/s~\cite{Fryer:2004}.
Multidimensional simulation by H.-T. Janka e.a.~\cite{Scheck:2006} where the explosion anisotropies develop chaotically, resulted in a possible pulsar velocity of $10^3$ km/s. 
However, there was no correlation between the direction of pulsar velocity and the rotation axis direction~\cite{Johnston:2005} in this approach.

Along with the hydrodynamic approach, there
were several different ideas to explain the velocities of pulsars, but all of them operated at speeds of scale of 100 km/s:
\begin{itemize}
\item[i)] 
evolution of close binary systems~\cite{Gott:1970};
\item[ii)] 
acceleration of a pulsar within a few months after the explosion due to asymmetric electromagnetic radiation
caused by the inclination of the magnetic moment with respect to the axis of rotation and the displacement of
the center of the star~\cite{Harrison:1975};
\item[iii)] 
asymmetric radiation of neutrinos (antineutrinos) in a collapse via the URCA-processes in a strong
magnetic field of the scale of $10^{14} - 10^{15} \, $ G 
in a supernova core~\cite{Chugai:1984,Loskutov:1984,Dorofeev:1984}. 
\end{itemize}

The neutrino mechanism looks the most interesting. It is known that neutrinos carry away 
99 \% of the total supernova energy $E \sim 3 \times 10^{53}\,$ erg. When the asymmetry is of $\sim 3\,$\%, neutrinos
carry the momentum of $\sim 0.03 \, E/c$. The compact explosion remnant, a neutron star with a mass 
$\sim 1.4 M_\odot$, gets the same momentum.  
In this case, its velocity can be easily estimated as $\sim$ 1000 km/sec.

However, neutrinos produced in the electroweak processes have small mean free paths in matter of the 
central part of a supernova and may not cause high-velocity pulsars~\cite{Vilenkin:1995,Lai:1998,Arras:1999}. 

A lively discussion was generated by the idea~\cite{Kusenko:1996}, under which the 
neutrino flux asymmetry from a protoneutron star arose due to neutrino oscillations in matter and intensive
magnetic field. The neutrinosphere for $\nu_\tau$ lies inside the neutrinosphere for $\nu_e$, and the resonant transition $\nu_e \to \nu_\tau$ is possible under certain conditions in the region between the neutrinospheres, 
where $\nu_e$ are entangled in the medium while $\nu_\tau$ are ``free'' to depart. Hence the surface of the resonant
transition becomes an effective neutrinosphere for $\nu_\tau$. In the presence of a magnetic field, 
this sphere is deformed along the field. Due to the temperature dependence on the radius, the anisotropy of the energy flux carried away by neutrinos, arises. This should cause the kick of the nascent neutron star.

The idea of the pulsar kick due to deformed neutrinosphere~\cite{Kusenko:1996}
raised, however, serious criticism~\cite{Janka:1998}: 
after the neutrinosphere deformation, the surfaces of the constant temperature would be deformed also, 
because just neutrinos provided a thermal equilibrium.
And the main problem of the model became clear soon: the existence of neutrinos with the mass $\sim$ 100 eV was needed.
Established restriction on the neutrino mass, $m_\nu < 2\,$ eV, ``closed'' the model.

There were also attempts to explain large space velocities of young pulsars with using 
of some possible non-standard properties of neutrinos. For example, a mechanism was proposes 
by E. Akhmedov e.a.~\cite{Akhmedov:1997}, of the resonant spin-flavour precession 
of neutrinos with a transition magnetic moment in the magnetic field of a supernova. 
The asymmetric emission of neutrinos was caused by the distortion of the resonance surface 
due to matter polarisation effects in the supernova magnetic field. 
The requisite values of the field strengths were declared to be of order $10^{16}\,$G, and neutrino parameters were taken within the existing experimental bounds. 
However, as was mentioned in the paper~\cite{Janka:1998}, in fact the magnetic fields were required 
in the model~\cite{Akhmedov:1997} more than an order of magnitude larger. 
 

\section{The initial pulsar kick and sterile neutrinos}


\indent\indent
The sterile neutrinos came on stage in Ref.~\cite{Kusenko:1997} (see also~\cite{Kusenko:2009} for details), where 
the same deformed by magnetic field neutrinosphere as in Ref.~\cite{Kusenko:1996} 
was discussed, but instead of oscillations
$\nu_{\mu,\tau} \leftrightarrow \nu_e$, the transitions were considered into ``heavy''
sterile neutrinos $\nu_{\mu,\tau} \leftrightarrow \nu_s$. 
The attractiveness of the model was in an idea that the heavy sterile neutrino (with the mass-scale of a 
few keV) simultaneously solved two problems: providing an initial velocity of pulsars
they could also play the role of dark matter.

However, when we have reproduced calculations performed in Refs.~\cite{Kusenko:1997,Kusenko:2009} 
we have obtained that the result for the asymmetry was overvalued in~\cite{Kusenko:2009} at 15 times. 
In other words, for the declared asymmetry, the necessary magnetic field strength should be 
15 times larger: not $\sim 3 \times 10^{16}\,$ G but $\sim 4.6 \times 10^{17}\,$ G.

Another scenario of using sterile neutrinos for the pulsar kick explanation 
with off-resonance transitions was developed in Ref.~\cite{Fuller:2003}. 
Sterile neutrinos could be born in $\beta$-processes due to the neutrino mixing, but the process 
is suppressed because of smallness of the mixing angle.
However, they can take a significant amount of energy due to two
factors:

(1) within the core, neutrinos have energies $\sim 150\,$ MeV,
which is much greater than the energy of active neutrinos $\sim 20\, $ MeV, emitted from a 
neutrinosphere;

(2) the emission occurs from the volume, not from the surface.

In the presence of a magnetic field, neutrinos are emitted asymmetrically, and this
asymmetry is maintained because sterile neutrinos are 
not absorbed, but fly away freely, unlike the situation in the approach of Refs.~\cite{Chugai:1984,Loskutov:1984,Dorofeev:1984}.
However, as our analysis shows, the asymmetry was overvalued in~\cite{Fuller:2003} at 40 times at least. 
In other words, for the declared in Ref.~\cite{Fuller:2003} asymmetry, the magnetic field strength 
needed should be 40 times larger: not $\sim 10^{16}\,$ G but $\sim 4 \times 10^{17}\,$ G.
In our opinion, the authors~\cite{Fuller:2003} made a mistake in calculation of the value $k_0$ 
defined in Eq.~(9) and presented in Fig.~2 of~\cite{Fuller:2003}. By the way, the authors call the value $k_0$ as 
the fraction of electrons in the lowest Landau level, while in fact it is 
the fraction of electron energy squared in the lowest Landau level, defining an asymmetry of 
the neutrino--electron interaction in $\beta$-processes. 
It can be shown that the result of the paper~\cite{Fuller:2003} is wrong, 
both by direct numerical calculation and analytically. 
Really, using Eqs.~(9), (10) of Ref.~\cite{Fuller:2003} one can transform  
the expression for the value $k_0$ with a good accuracy to the form:
\begin{equation}
k_0 \simeq \frac{e B}{2 T^2} \; \frac{J_2 (\mu_e/T)}{J_4 (\mu_e/T)} \,,
\label{eq:k_0}
\end{equation}
where $B$ is the magnetic field strength, $\mu_e$ and $T$ are the chemical potential and the temperature 
of electrons, and $J_n (\eta)$ are the Fermi integrals:
\begin{equation}
J_n (\eta) = \int\limits_0^\infty \frac{x^n \, \D x}{\E^{x-\eta} + 1} \,.
\label{eq:J_n}
\end{equation}
Depending on the electron chemical potential and the magnetic field strength, the value $k_0$ 
is overestimated in Fig.~2 of~\cite{Fuller:2003} by the factor from 40 to 90. 

In the recent e-print~\cite{Kishimoto:2011}, C. Kishimoto presented 
a detailed numerical analysis of 
the transformation of active neutrinos to sterile neutrinos through an MSW-like resonance 
in the protoneutron star, in order to provide the pulsar kick. 
However, after correcting a numerical mistake in the version 1 of the e-print, it can be seen 
that the magnetic field strength needed for a desirable effect should be taken 
of the order of $10^{18}\,$ G.


\section{Back to standard neutrinos?}


\indent\indent
A reasonable question arises: if we really need such strong magnetic fields to provide 
a natal neutron star kick with sterile neutrinos, 
isn't it possible to manage with standard neutrinos?

As it was already mentioned, an asymmetry of the standard neutrino radiation
in a strong magnetic field is not a new topic. 
For example, in the series of papers by our 
group~\cite{Kuznetsov:1997,Kuznetsov:1999,Gvozdev:1999}
the asymmetry of neutrino emission in a strong magnetic field:
\begin{equation}
A \; = \; \frac{|\sum_i {\bf p}_i|}{\sum_i |{\bf p}_i|} \,
\label{eq:asymm1}
\end{equation}
was analysed, which arised due to parity violation in the neutrino-electron 
and neutrino-nucleon processes.
In a strong field of the poloidal type~\cite{Duncan:1992,Bocquet:1995,Cardall:2001}, 
only due to the process $\nu \to \nu e^- e^+$ one obtains~\cite{Kuznetsov:1997}:
\begin{equation}
A  \sim  3 \times 10^{-3}\, 
\left (\frac{B}{10^{16}~\mbox{G}} \right )
\left (\frac{\bar E}{20 ~\mbox{MeV}}\right )^3 
\left (\frac{ \Delta \ell}{20 ~\mbox{km}}\right ),
\label{eq:asymm2}
\end{equation}
where $\Delta \ell$ is the characteristic size of the region 
where the field strength varies insignificantly, and
$\bar E$ is the neutrino energy averaged over the neutrino spectrum.
The asymmetry is seen to be not enough to provide the 
observable neutron star kick with such field strength. 

It should be noted that the mechanism is known of essential enhancement 
of the magnetic field strength during a supernova explosion. 
It is the magnetorotational supernova by G.S.~Bisnovatyi-Kogan~\cite{Bisnovatyi-Kogan:1970,Ardeljan:2005}, 
a model for generation of the toroidal magnetic field. 
A poloidal magnetic field being enhanced during the supernova core collapse and being frozen in plasma, due to the differential rotation, generates a strong toroidal magnetic field which could be in order of 
magnitude greater than the original poloidal field.


\section{Tangential Neutrino Force}


\indent\indent
Let us remind what is a possible integral effect of neutrinos on a magnetized plasma. 
Consider first the neutrino-electron processes~\cite{Kuznetsov:1999}. 
A complete set of these processes in plasma, 
\begin{equation}
\nu e^\mp \to \nu e^\mp, \quad \nu \to \nu e^- e^+, \quad \nu e^- e^+ \to \nu \,,
\label{eq:processes}
\end{equation}
lead to the energy and force neutrino flux impact on plasma:
\begin{equation}
(\dot{\cal E}, {{\cal F}_z}) 
= \int (P-P')_{0,{z}} \; \D n_\nu \, \D W \, , \qquad
\D n_\nu = \frac{\D^3 P}{(2 \pi)^3} \,\,\frac{\Phi (\vartheta, R)}{{\E^{(E - \mu_\nu)/T_\nu} + 1}} \, .
\label{eq:force1}
\end{equation}
Here, 
$\D W$ is the total differential probability of all the processes specified in~\eqref{eq:processes},
$P$ and $P'$ are the initial and final neutrino four-momenta, 
the $z$ axis is directed along the magnetic field, 
$\D n_\nu$ is the initial neutrino density, 
$\mu_\nu$ and $T_\nu$ are the effective chemical potential and the spectral temperature of the neutrino gas, 
and the function $\Phi (\vartheta, R)$ determines the neutrino angular distribution, 
depending on the angle $\vartheta$ between the neutrino momentum and the radial direction 
in the star and on the distance $R$ from the center of the star. 
It should be noted that Eq.~\eqref{eq:force1} can be used for evaluating the integral effect of neutrinos on plasma 
in the conditions of not very dense plasma, e.g. of a supernova envelope, 
when an one-interaction approximation of a neutrino with plasma is valid. 

Spectral temperatures for different types of neutrinos are estimated to be~\cite{Imshennik:1989}:
\begin{equation}
T_{\nu_e} \simeq 4 \,\mbox{MeV}, \quad
T_{\bar\nu_e} \simeq 5 \,\mbox{MeV}, \quad
T_{\nu_{\mu,\tau}} \simeq T_{\bar\nu_{\mu,\tau}} \simeq 8 \,\mbox{MeV} .
\label{eq:T}
\end{equation}
The probability of the $\beta$ processes ($\nu_e + n \leftrightarrow e^- + p$) 
is substantially higher than that for neutrino-electron processes, so the $\beta$ processes
dominate in the energy balance. As a result of neutrino heating the plasma, temperature should be 
very close to the spectral temperature of the electron neutrinos, $T \simeq T_{\nu_e}$. 

As it was shown in Refs.~\cite{Kuznetsov:1999}, the main contributions into the values 
$\dot{\cal E}$ and ${\cal F}_z$ in~\eqref{eq:force1} were made by $\mu$ and $\tau$ neutrinos and 
antineutrinos (as a result of the conservation of $CP$, neutrinos and antineutrinos
push the plasma in the same direction). This is because in the vicinity of the $\nu_e$ 
neutrinosphere the spectral temperatures of the other types of neutrinos differ substantially
from the plasma temperature $T \simeq T_{\nu_e}$. 

For numerical estimates we can conveniently express the contribution from $\nu$-$e$-processes
with $\bar\nu_e, \, \nu_{\mu,\tau}, \, \bar\nu_{\mu, \tau}$ into 
the values $\dot{\cal E}$ and ${\cal F}_z$ in the following form: 
\begin{equation}
(\dot{\cal E}, {\cal F}_z)_{\nu_i} 
\simeq {\cal A}\;
\left[\left(C_V^{(i)}\right)^2 + \left(C_A^{(i)}\right)^2, {2 C_V^{(i)} C_A^{(i)}} \right] \, \psi (T_{\nu_i}/T) ,
\label{eq:force2}
\end{equation}
where 
\begin{equation}
{\cal A} = \frac{12\, G_{\mathrm{F}}^2 \,e B \,T^7}{\pi^5} =
\left(\frac{B}{{10^{16} \mbox{G}}}\right)
\left (\frac{T}{{4\;\mbox{MeV}}} \right )^7 \times 
\begin{cases}
1.6 \cdot 10^{30}\;\frac{\mbox{erg}}{\mbox{cm}^3 \cdot\mbox{s}},\cr \cr
0.55 \cdot 10^{20}\;\frac{\mbox{dyne}}{\mbox{cm}^3},\cr
\end{cases}
\label{eq:A}
\end{equation}
The electroweak constants $C_V^{(i)}, C_A^{(i)}$ of the effective Lagrangian of the neutrino-electron interaction 
in Eq.~\eqref{eq:force2} are:
\begin{equation}
C_V^{(e)} = \frac{1}{2} + 2 \sin^2 \theta_{\rm W} \,, \; C_A^{(e)} = \frac{1}{2} \,, \qquad 
C_V^{(\mu,\tau)} = - \frac{1}{2} + 2 \sin^2 \theta_{\rm W} \,, \; C_A^{(\mu,\tau)} = - \frac{1}{2} \,.
\label{eq:CV_CA}
\end{equation}
The temperature dependent function has the form:
\begin{equation}
\psi (\tau_i) = \frac{\tau_i^7}{6} \, \int\limits_0^\infty 
\frac{y^2 \D y}{\E^{\tau_i y} - 1} \left [\E^{(\tau_i - 1) y} - 1 \right ], \quad
\psi (\tau_i)\bigg\vert_{\tau_i \to 1} \simeq \frac{\pi^4}{90} \, (\tau_i - 1) \,.
\label{eq:psi}
\end{equation}
For electron antineutrinos one obtains $\psi (1.25) \simeq 0.82$, while 
for muon and tau neutrinos and antineutrinos the function is
$\psi (2) \simeq 38.5$.

A combined effect of all types of neutrinos interacting with electron-positron plasma is:
\begin{equation}
{\cal F}_B^{(\nu e)} \simeq 
3.6 \times 10^{20} 
\left(\frac{B}{{10^{16} \mbox{G}}}\right) 
\left(\frac{T}{{4\;\mbox{MeV}}} \right )^7
\;\frac{\mbox{dyne}}{\mbox{cm}^3}\,.
\label{eq:force_nu_e}
\end{equation}

Contribution of the neutrino-nucleon processes was evaluated in Refs.~\cite{Gvozdev:1999}. 
For the parameters of the shell of a supernova: $Y_e \simeq 0.2, \; 
\rho \simeq 10^{11-12} \,$ g/cm$^3$, one obtains (`$\nu N$' means both urca-processes and $\nu N$-scattering)
\begin{equation}
{\cal F}_B^{(\nu N)} \simeq 
2.4 \times 10^{20} 
\left(\frac{B}{{10^{16} \mbox{G}}}\right) 
\;\frac{\mbox{dyne}}{\mbox{cm}^3}\,.
\label{eq:force_nu_N}
\end{equation}
It is important that the contributions of both neutrino-electron and neutrino-nucleon processes 
are of the same sign. The total neutrino force density is:
\begin{equation}
{\cal F}_B^{(total)} \simeq 
0.6 \times 10^{21} 
\left(\frac{B}{{10^{16} \mbox{G}}}\right) 
\;\frac{\mbox{dyne}}{\mbox{cm}^3}\,.
\label{eq:force_nu_tot}
\end{equation}

Note that the force density~\eqref{eq:force_nu_tot} is five orders of
magnitude lower than the density of the gravitational
force and thus negligibly influences the radial dynamics
of the supernova shell. However, when a toroidal
magnetic field~\cite{Bisnovatyi-Kogan:1970,Ardeljan:2005} is generated in the shell, 
the force~\eqref{eq:force_nu_tot} 
directed along the field can fairly rapidly (within
times of the order of a second~\footnote{We know that the cooling stage of a supernova shell, 
known as the Kelvin--Helmholz stage, lasts for around 10 s.}) 
lead to substantial redistribution
of the tangential plasma velocities. Then in
two toroids in which the magnetic field has opposite
directions, the tangential neutrino acceleration of the
plasma will have different signs relative to the rotational
motion of the plasma. This effect can then lead to
substantial redistribution of the magnetic field lines,
concentrating them predominantly in one of the toroids.
This leads to considerable asymmetry of the magnetic
field energy in the two hemispheres and may be
responsible for the asymmetric explosion of the supernova
which could explain the discussed phenomenon of high
intrinsic pulsar velocities. In our view it is interesting
to model the mechanism for toroidal magnetic field
generation taking into account the neutrino force action
on the plasma both via neutrino--nucleon and neutrino--electron processes.


\section{Neutrino-triggered magnetorotational pulsar natal kick}


\indent\indent
Considered neutrino processes in toroidal magnetic field which is frozen
in plasma, provide the angular acceleration for an element of plasma at a distance $R$ from the rotation axis:
\begin{equation}
\dot\Omega = \frac{\cal F}{\rho \, R} \simeq 1.2 \times 10^3 
\frac{1}{\mbox{sec}^2}
\left(\frac{B}{{10^{16} \mbox{G}}}\right) .
\label{eq:ang_acc}
\end{equation}
It means that during the time $\sim$ 1 sec the increase of the angular velocity is
\begin{equation}
\Delta \Omega \sim 10^3 
\frac{1}{\mbox{sec}}
\left(\frac{B}{{10^{16} \mbox{G}}}\right) .
\label{eq:ang_vel}
\end{equation}
In the one hemisphere, the angular acceleration coincides with the direction
of the initial rotation, while in another hemisphere, they are opposites.
A neutrino flux, pushing the plasma, torques the toroids in different directions. 

Thus, three stages of a pulsar kick can be identified:  
\begin{itemize}
\item[i)] 
pre-supernova core is collapsing with rotation during 0.1 sec when a strong toroidal magnetic field is 
generated due to the differential rotation; 
\item[ii)] 
the neutrino outburst, pushing the plasma by the tangential force along the toroidal magnetic field which is frozen in plasma, leads to a~magnetic field asymmetry: the field strength is enhancing in one hemisphere and is decreesing in another one, during $\sim$ 1 sec;  
\item[iii)] 
the pressure difference arising in the two hemispheres, causes the kick to a core. 
\end{itemize}

According to the momentum conservation, 
an energetic plasma jet must be formed opposite to the pulsar velocity 
direction~\footnote{This remark was made by Hans-Thomas Janka.}. 

Surely, a detailed multy-dimensional numerical simulation of the process is needed. 
Let us estimate in order of magnitude what to expect.

A pressure difference arising in the two hemispheres can be evaluated as:
\begin{equation}
\Delta p \simeq \frac{B^2}{8 \pi} = \frac{(e B)^2}{8 \pi \alpha} \,, 
\label{eq:pressure}
\end{equation}
where $\alpha = 1/137$ is the fine-structure constant. 
The magnetic field pressure causes the plasma acceleration:
\begin{equation}
\frac{\D V_{kick}}{\D t} \simeq 1.6 \times 10^5 \, \frac{\mbox{km}}{\mbox{sec}^2}
\left(\frac{B}{{10^{16} \mbox{G}}}\right)^2 
\left(\frac{R}{{20 \,\mbox{km}}}\right)^2 
\left(\frac{1.4 \, M_\odot}{M}\right)
\sin 2\theta \; \Delta \theta 
 \,, 
\label{eq:acc_1}
\end{equation}
where $R, \theta$ and $\Delta \theta$ are the parameters that characterize the region 
of a strong toroidal magnetic field, see Fig.~1. 

\begin{figure}
\begin{center}
\includegraphics*[width=0.5\textwidth]{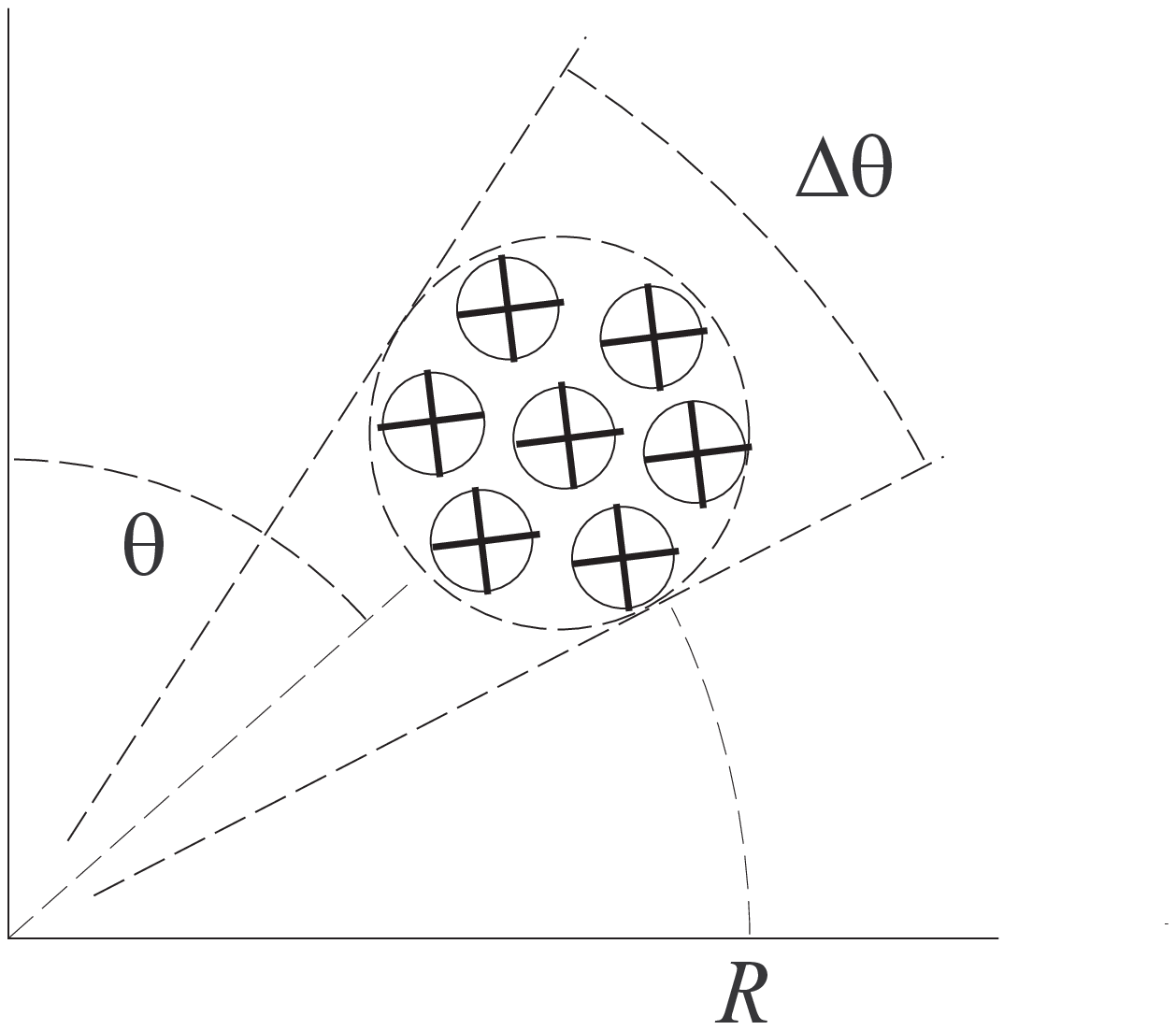}
\caption{The region of a strong toroidal magnetic field.} 
\end{center}
\label{fig:1}
\end{figure}

Taking for estimation $\Delta \theta \sim 15^0 \sim \frac{1}{4} \, $, $\theta \sim 45^0$ 
one obtains
\begin{equation}
\frac{\D V_{kick}}{\D t} 
\simeq 4 \times 10^4 \, \frac{\mbox{km}}{\mbox{sec}^2}
\left(\frac{B}{{10^{16} \mbox{G}}}\right)^2 
\left(\frac{R}{{20 \,\mbox{km}}}\right)^2 
\left(\frac{1.4 \, M_\odot}{M}\right)
 \,. 
\label{eq:acc_2}
\end{equation}

In fact, the acceleration is not a constant, and an expansion of the magnetic field volume, 
which reduces the magnitude of the field should be taken into account. 
The magnetic flux conservation provides: $p \,V^2 = \mbox{const}$. 

Within the same geometry, one obtains:
\begin{equation}
V_{kick} \simeq 600 \, \frac{\mbox{km}}{\mbox{sec}}
\left(\frac{B_0}{{10^{16} \mbox{G}}}\right) 
\left(\frac{R}{{20 \,\mbox{km}}}\right) 
\left(\frac{\Delta \, z}{{5 \,\mbox{km}}}\right)^{1/2} 
\left(\frac{1.4 \, M_\odot}{M}\right)^{1/2}
 \,, 
\label{eq:kick}
\end{equation}
where $B_0$ is the initial field strength, $\Delta \, z$ is a distance traveled 
by a compact remnant of the explosion. 

\section{Conclusions}


\begin{itemize}
\item
There are many mechanisms for pulsar natal kick, and the one using sterile neutrinos 
and proposed by A.~Kusenko e.a. looks the most attractive. However, as the analysis shows, 
for the declared effect the magnetic field strength should be 
much larger, not $\sim 10^{16}\,$ G, but $\gtrsim 4 \times 10^{17}\,$ G.

\item
With such strong magnetic fields, it is possible to manage with standard neutrinos. 

\item
As it was shown in the papers by our group, neutrino-electron and neutrino-nucleon processes 
in a strong magnetic field, cause the appearance of a force density acting on magnetized plasma along the field direction,
$$ 
{\cal F}_B \simeq 
0.6 \times 10^{21} 
\left(\frac{B}{{10^{16} \mbox{G}}}\right) 
\;\frac{\mbox{dyne}}{\mbox{cm}^3}\,.
$$

\item
If the strong toroidal magnetic field is generated in the vicinity of the supernova core (magnetorotational supernova
model by G.S.~Bisnovatyi-Kogan), the neutrino flux, pushing the plasma, torques the toroids in different directions. 
In the one hemisphere, the additional angular acceleration coincides with the direction
of the initial rotation, while in another hemisphere, they are opposites.

We stress that just the toroidal magnetic fields are considered which can be generated in order of 
magnitude greater than the poloidal fields used in other approaches. 

\item
There arises the magnetic field asymmetry in the two hemispheres, and consequently the field 
pressure difference providing the pulsar kick acceleration:
$$
\frac{\D V_{kick}}{\D t} 
\simeq 4 \times 10^4 \, \frac{\mbox{km}}{\mbox{sec}^2}
\left(\frac{B}{{10^{16} \mbox{G}}}\right)^2 
\left(\frac{R}{{20 \,\mbox{km}}}\right)^2 
\left(\frac{1.4 \, M_\odot}{M}\right)
$$

and the kick velocity:

$$
V_{kick} \simeq 600 \, \frac{\mbox{km}}{\mbox{sec}}
\left(\frac{B_0}{{10^{16} \mbox{G}}}\right) 
\left(\frac{R}{{20 \,\mbox{km}}}\right) 
\left(\frac{\Delta \, z}{{5 \,\mbox{km}}}\right)^{1/2} 
\left(\frac{1.4 \, M_\odot}{M}\right)^{1/2}
$$

\item
Because of large acceleration, a pulsar acquires big velocity during very short time, 
like in a shot. 
One should remember how he operated, being a child, with a cherry-stone after eating cherries: 
pressing it asymmetrically by fingers, he provided a big velocity to that compact object. 
So, we may consider a kind of ``Cherry-Stone Shooting'' mechanism for pulsar natal kick.

\item
According to the momentum conservation, 
an energetic plasma jet must be formed opposite to the pulsar velocity 
direction.

\item
A detailed multy-dimensional numerical simulation of the process is needed. 
We would believe it should confirm the effect. 
\end{itemize}


\section*{Acknowledgements}


\indent\indent
We are grateful to V. A. Rubakov and all the participants of the seminar 
at the Theoretical Physics Department of the Institute for Nuclear Research 
of the Russian Academy of Sciences for useful discussion.  
We thank H.-T. Janka and A. V. Borisov for interesting remarks and 
C.~Kishimoto for correspondence.  
A. K. expresses his deep gratitude to the organizers of the 
XV-th International School ``Particles and Cosmology'' 
for warm hospitality.

This work was performed in the framework of realization of the Federal
Target Program ``Scientific and Pedagogic Personnel of the Innovation
Russia'' for 2009 - 2013 (State contract no. P2323)
and was supported in part by the Ministry of Education
and Science of the Russian Federation under the Program
``Development of the Scientific Potential of the Higher
Education'' (project no. 2.1.1/13011), and by the Russian Foundation
for Basic Research (project no. 11-02-00394-a).



\end{document}